\documentclass{elsart}
\usepackage{ifpdf}
\usepackage{graphicx,natbib,amssymb,lineno}
\ifpdf
\journal{Planetary and Space Science}
\usepackage[
  pdftitle={Instructions for use of the document class
    elsart},%
  pdfauthor={Simon Pepping},%
  pdfsubject={The preprint document class elsart},%
  pdfkeywords={instructions for use, elsart, document class},%
  pdfstartview=FitH,%
  bookmarks=true,%
  bookmarksopen=true,%
  breaklinks=true,%
  colorlinks=true,%
  linkcolor=blue,anchorcolor=blue,%
  citecolor=blue,filecolor=blue,%
  menucolor=blue,pagecolor=blue,%
  urlcolor=blue]{hyperref}
\usepackage[%
  breaklinks=true,%
  colorlinks=true,%
  linkcolor=blue,anchorcolor=blue,%
  citecolor=blue,filecolor=blue,%
  menucolor=blue,pagecolor=blue,%
  urlcolor=blue]{hyperref}
\fi
\newcommand{\Msun}{\mbox{\rm M$_{\odot}$}}

\makeatletter
\def\elsartstyle{%
    \def\normalsize{\@setfontsize\normalsize\@xiipt{14.5}}
    \newcommand{\Msun}{\mbox{\rm M$_{\odot}$}}
    \def\small{\@setfontsize\small\@xipt{13.6}}
    \let\footnotesize=\small
    \def\large{\@setfontsize\large\@xivpt{18}}
    \def\Large{\@setfontsize\Large\@xviipt{22}}
    \skip\@mpfootins = 18\p@ \@plus 2\p@
    \normalsize
}
\@ifundefined{square}{}{}
\makeatother

\begin{document}

\begin{frontmatter}
\title{Constraints from deuterium on the formation of icy bodies in the Jovian system and beyond}

\author{Jonathan Horner$^{1}$}
\ead{j.a.horner@open.ac.uk}
\author{Olivier Mousis$^{2}$,}
\author{Yann Alibert$^{2}$,}
\author{Jonathan I. Lunine$^{3}$}
\and
\author{Michel Blanc$^{4,5}$}
\address{
$^1$ Astronomy Group, The Open University, Walton Hall, Milton Keynes, 
MK7 6AA, UK\\
$^2$Observatoire de Besan\c{c}on, Institut UTINAM, CNRS-UMR 6213, BP 1615, 25010 Besan\c{c}on Cedex, France\\
$^3$Lunar and Planetary Laboratory, University of Arizona, Tucson, AZ, USA\\
$^4$Universit{\'e} Paul Sabatier, Centre d'Etude Spatiale des Rayonnements, France\\
$^5$Ecole Polytechnique Palaiseau Cedex, France\\}

\begin{abstract}
We consider the role of deuterium as a potential marker of location
and ambient conditions during the formation of small bodies in our
Solar system. We concentrate in particular on the formation of the
regular icy satellites of Jupiter and the other giant planets, but
include a discussion of the implications for the Trojan
asteroids and the irregular satellites.  We examine in detail the
formation of regular planetary satellites within the paradigm of a
circum-Jovian subnebula. Particular attention is paid to the two
extreme potential subnebulae -- ``hot'' and ``cold''. In particular, we
show that, for the case of the "hot" subnebula model, the D:H ratio in
water ice measured from the regular satellites would be expected to be
near-Solar. In contrast, satellites which formed in a ``cold'' subnebula
would be expected to display a D:H ratio that is distinctly
over-Solar.  We then compare the results obtained with the enrichment
regimes which could be expected for other families of icy small bodies
in the outer Solar system -- the Trojan asteroids and the
irregular satellites. In doing so, we demonstrate how measurements by
Laplace, the James Webb Space Telescope, HERSCHEL and ALMA will play an important 
role in determining the true formation
locations and mechanisms of these objects.
\end{abstract}

\end{frontmatter}

\section{Introduction}
\label{intro}

The physical and chemical characteristics of the proto-planetary
nebula from which our solar system formed can be inferred through the
analysis of ``primitive'' objects such as meteorites, comets, and the
giant planets themselves. We can obtain useful constraints on these
processes by examining the degree to which fossil deuterium contained
within the water in some of these objects is enriched when compared to
the protosolar abundance, which can be measured in various objects
within the Solar system. Within the Solar nebula, the main reservoir
of deuterium was molecular hydrogen. However, isotopic exchange
occurred between this hydrogen and other deuterated species, resulting
in the formation of secondary reservoirs of deuterium. As a result of
the high cosmic abundance of oxygen, the most important secondary
reservoir in the nebula is water (HDO), either in gaseous or solid
phase.  

Calculations of the temporal and radial evolution of the D:H ratio in
the primitive nebula (Drouart et al., 1999; Mousis et al., 2000) have
been performed in order to reproduce existing data on comets (Balsiger
et al., 1995, Eberhardt et al., 1995; Bockel{\'e}e-Morvan et al.,
1998; Meier et al., 1998; see Horner et al., 2007), and measurements
taken from meteorites (Deloule et al., 1998). One particularly
interesting result of these calculations is that the D:H ratio in
water ice produced in the nebula varies by a significant amount as a
function of the distance from the Sun at which the ice was formed.
This variation can be seen clearly in the compilation of
  measurements given by Drouart et al. (1999), in Fig.1 of that work.
Such results led Horner et al. (2007) to discuss how measurements of
the D:H ratio in cometary bodies might prove helpful in answering the
question of whereabouts in the Solar system the different cometary
populations had formed. The study of D:H extends beyond the study of
cometary bodies, however.

In order to consider the measured D:H enrichment within an object to
be the direct result of its formation, it is clear that the deuterated
reservoir incorporated within that object must have undergone little
or no alteration since its formation within the nebula. Therefore,
despite the fact that deuterium enhancement has been measured on the
Earth, Venus and Mars, it is obvious that we cannot consider
the resulting value to be primordial. Indeed, the value measured on
the Earth could be the result of the combination of volatile material
from a number of sources (Dauphas et al., 2000). Futhermore, the
values measured on Venus and Mars are believed to be the result of
strong atmospheric fractionation, which has occurred throughout the
history of the planets (Donahue, 1999; Bertaux \& Montmessin, 2001).

Even further from the Sun, observations of Titan reveal an
unexpectedly high D:H ratio within the methane of the satellite's
atmosphere (B{\'e}zard et al., 2007).  The cause of this enhancement
is still under some debate. It could be the result of isotopic
exchange between methane and molecular hydrogen within the early Solar
nebula, prior the formation of the icy planetesimals that were
ultimately accreted by Titan (e.g. Mousis et al., 2002a).
Alternatively, it could result from a photochemical process occurring
within the atmosphere of the satellite (Lunine et al., 1999).

 A number of ambitious projects, such as the proposed
  Laplace\footnote{http://jupiter-europa.cesr.fr/} mission and the
  James Webb Space Telescope (scheduled for launch in 2013), will
  allow us to revisit the Jovian satellite system, and provide new
  measurements of the satellites of the giant planet in the coming
  years. As such, the time seems right to revisit the question of how
  these satellites formed, and what effect their formation would have
  on the quantities that could be observed by such a mission. In light
  of these proposed missions, such work takes on an interesting new
  aspect. How would the D:H ratio in the Jovian (or Saturnian)
  satellites be affected by their formation?  Could measurements of
  deuterium in these satellites help us to understand the final stages
  of the formation of their parent bodies?  Furthermore, it is
  important to examine such ideas during the period over which the
  missions are designed, to enable the construction of instruments
  which are fully capable of answering the questions posed.

In this work, we aim to detail the various properties which would have
affected the regular satellites during their formation, highlighting
how the fractionation of deuterium within their ices may represent a
vital window into their formation. We also examine the benefits that
observations of the degree of deuteration in the irregular satellites
and the Trojan asteroids would have for our understanding of
the origin of the volatiles they contain, together with presenting a
discussion of how and when such observations may be made.

The simple picture portrayed for the comets, although applicable for
other objects which formed freely in the Solar nebula (such as the
asteroids), becomes more complicated when one wishes to understand the
formation of the satellites of the giant planets. For the regular
satellites, it is possible that material from the Solar nebula
underwent a significant amount of additional processing within the
planetary sub-nebula. In section 2, we will review the main results of
prior discussions and calculations involving the D:H ratio in the
Solar nebula, while in section 3 we show how further processing within
planetary sub-nebulae would eventually lead to a reduction in the D:H
ratio incorporated in the satellites forming therein when compared to
 gas at an equivalent distance in the Solar nebula. In
section 4, we examine the cases of the irregular planetary satellites
and the Jovian Trojan asteroids, two groups of object for which the
study of deuteration may help untangle the minutae of their
formation. The measurement of the D:H ratio in satellites and other
objects by future space missions is discussed in section 5, and, in
section 6, we conclude with a summary and discussion of our ideas,
along with their implications for future missions to, and observations
of, the giant planets and their satellite systems.

\section{Temporal and Radial evolution of D:H in the Solar nebula -- previous work}
\label{background}

Here, we summarise the works of Drouart et al. (1999), Mousis
et al. (2000), Hersant et al. (2001), Mousis (2004a) and Horner et
al. (2007) who described the evolution of the deuterium enrichment
factor, $f$, in H$_2$O within the Solar nebula. The calculations of
these authors were based on the fact that the main reservoir of
deuterium in the Solar nebula was molecular hydrogen (HD vs. H$_2$),
and that ion-molecule reactions in the interstellar medium (see
e.g. Brown \& Millar, 1989) resulted in the fractionation of deuterium
between deuterated species. Consequently, in the pre-solar cloud, such
fractionation was present, resulting in heavier molecules being
enriched. Water was the second most abundant hydrogen bearer in the
Solar nebula, as it is in our current Solar system, and therefore
became the second largest deuterium reservoir.

In the Solar nebula, the isotopic fractionation of deuterium between
water and hydrogen followed the reversible reaction (Geiss \& Reeves,
1981):

\begin{equation} 
\mathrm{H_2O + HD} \rightleftharpoons \mathrm{HDO + H_2} 
\end{equation}

At low temperatures, this reaction favors the concentration of
deuterium in HDO, but the reaction kinetics at such temperatures tend
to inhibit the enrichment of deuterium in water. The
enrichment factor, $f$, which results from the exchange between HD and
HDO is defined as the ratio of D:H in the considered deuterated
species to that in molecular hydrogen (the protosolar value). As a
result, for water we have:
 
\begin{equation} 
f =  {\mathrm{HDO / H_2O} \over \mathrm{HD / H_2} } 
\end{equation}

The afore-mentioned authors physically interpreted the measurements of
the D:H ratio in the LL3 meteorites, along with measurements taken of
comets 1P/Halley, C/1996 B2 (Hyakutake), and C/1995 O1 (Hale-Bopp)
(all three of which share a similar value of D:H in H$_2$O). Time
dependent turbulent models of the Solar nebula were then applied which
depend on three physical parameters: the initial mass of the nebula
M$_{D0}$, its initial radius R$_{D0}$, and the viscosity parameter
$\alpha$ (derived from the prescription of Shakura \& Sunyaev,
(1973)). They calculated $f$ in water with respect to the protosolar
value in molecular hydrogen by integrating an equation of diffusion in
the Solar nebula, as a function of the heliocentric distance and
time. The comparison of the obtained value of $f$ to observations
allowed a range of possible values for M$_{D0}$, R$_{D0}$, and
$\alpha$ to be determined.

This diffusion equation takes into account the isotopic exchange
between HDO and H$_2$ in the vapor phase, and turbulent diffusion
throughout the Solar nebula (see Sec. \ref{regular} for details). The
isotopic exchange between HDO and H$_2$ occurs as long as H$_2$O
remains in the vapor phase. This implies that the value of the
enrichment in microscopic ices is the one fixed at the time and at the
location of the condensation of the water vapor. As soon as the grains
reach millimetre size, they begin to decouple from the gas as they
grow further, leading to the formation of planetesimals. Whatever the
subsequent evolution of these bodies, their D:H ratio is that of the
microscopic grains from which they formed. In this paper, we consider
the case where the solids that grew in the Solar nebula were accreted
only from icy grains formed locally. This means that the D:H ratio in
the deuterated ices within these planetesimals is that which was
present at the time and location at which they condensed.

\begin{figure} 
\resizebox{\hsize}{!}{\includegraphics[angle=-90]{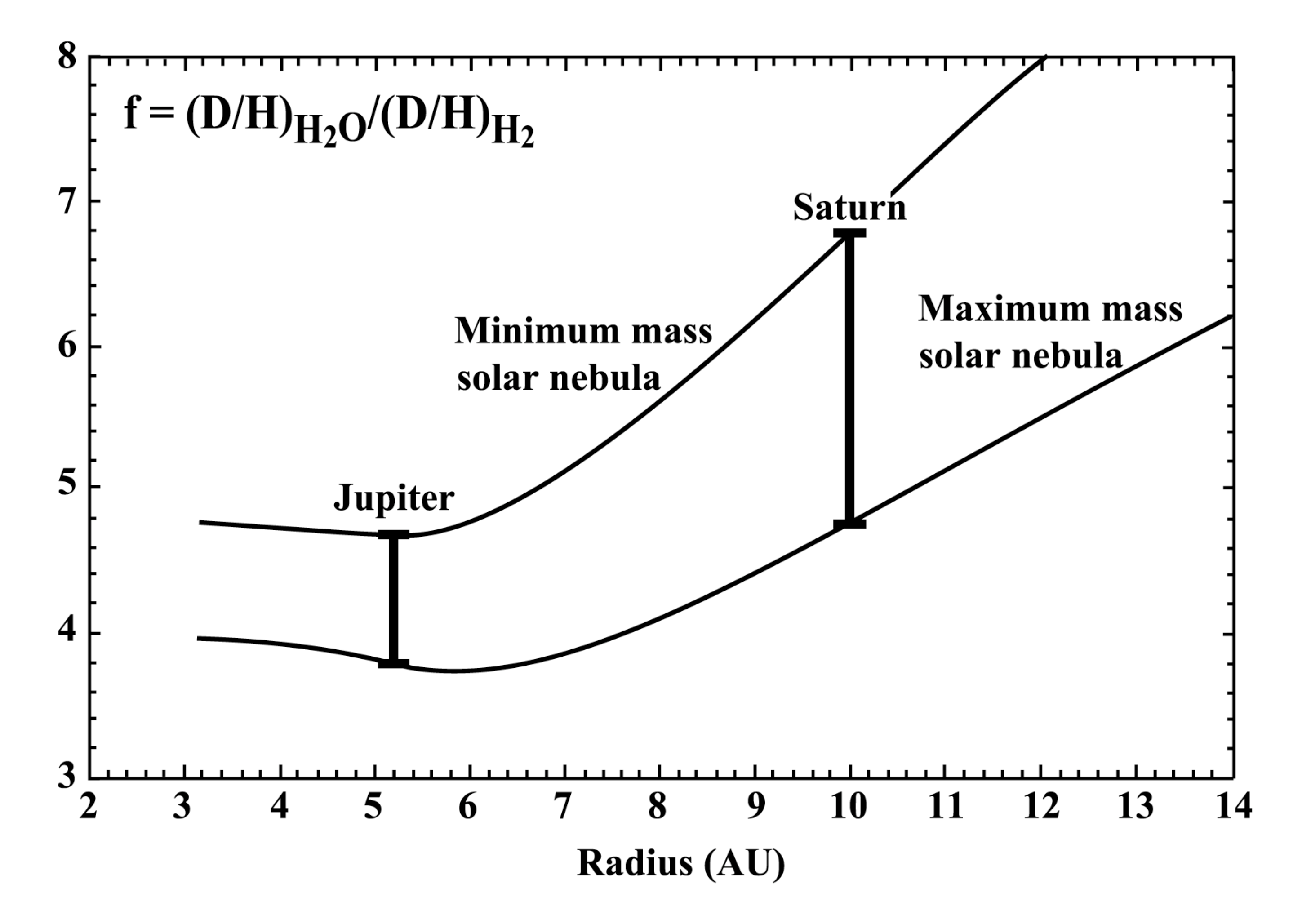}}
\caption{Deuterium enrichment factor $f$ calculated at the epochs of the condensation of water for the minimum mass and maximum mass solar nebula models (see e.g. Mousis 2004). The vertical bold solid lines correspond to the range of values of $f$ inferred in the icy solids produced in Jupiter and Saturn's feeding zones.}
\label{DHsol} 
\end{figure}

Figure \ref{DHsol} shows the variation of $f$ for water trapped within
icy grains, as a function of their distance from the Sun, in the case
of the minimum-mass and maximum-mass models of the Solar nebula
employed by Mousis (2004a) (see Table \ref{paraneb}). Here, as in
previous work, we assume that $f(R)$ = 31 at $t$ = 0 for D:H in
water. This value corresponds to that measured in the highly enriched
component (where D:H~=~$(73~\pm~12)~\times~10^{-5}$) found in LL3
meteorites (Deloule et al., 1998) compared to a protosolar value
assumed to be $(2.35 \pm 0.3) \times 10^{-5}$ (Mousis et al., 2002).

As discussed by Mousis (2004a), assuming that Jupiter and Saturn were
formed at their current locations in the Solar system (Pollack et
al., 1996), the plausible range of values of $f$ in the solids produced
in the feeding zones of both giant planets is covered by the vertical
error bars in Fig.\ref{DHsol}. The value of $f$ in H$_2$O in ices
produced in the feeding zone zone of Jupiter is thus between 3.8 and
4.7 times the Solar value. In the case of ices produced in the feeding
zone of Saturn, the value of $f$ in H$_2$O is ranging between 4.8 and
6.8 times the Solar value. Note that these calculations do not
consider the formation of planetesimals from a mixing of grains coming
from all over the nebula, due to turbulent mixing. In such a case, the
global enrichment factor in the inner regions should be somewhat higher,
while it would be expected to be a little lower in the outer regions, 
as the mixing ``smears" the distribution. This effect is discussed in 
more detail in Horner et al. (2007), and will not be considered further
in this work.

\section{Deuterium enrichment in the regular icy satellites}
\label{regular}

In the core-accretion model, the formation process of giant planets
can be separated in two phases, depending on the physical phenomenon
governing the rate of gas accretion. Indeed, during the first epoch of
formation, the large cooling timescale (which is necessary to allow
the release of the accretion energy of the planetesimals that build
the core) prevents the accretion of substantial amounts of gas. The
total radius of the planet is therefore equal to its Hill radius (the
region over which the planet's gravity exerts a stronger pull on
objects than that of the Sun). Once the planet reaches a certain mass,
the second epoch of its formation begins.  At this point, the cooling
timescale becomes low enough to allow a far greater gas accretion
rate, which may even be larger than that which can be supported by the
protoplanetary disk\footnote{ The maximum accretion rate that can be
supplied by the disk is governed by its surface density and
viscosity.}. At this stage, the total radius of the planet shrinks,
and it becomes significantly smaller than its Hill radius. Matter
flowing from the Solar nebula to the forming planet forms a
circum-planetary disk, known as a subnebula, in which regular
satellites are believed to form (Lubow et al., 1999; Magni and
Coradini, 2004).

The structure and evolution of the subnebula is therefore linked to
both the formation of the giant planet and the behaviour of the Solar
nebula, which supplies the gas and gas-coupled solids that may
ultimately take part in the formation of satellites (Canup \& Ward,
2002 \& 2006). The calculation of the structure and evolution of the
subnebula has been the subject of much scrutiny (see e.g.  Canup \&
Ward, 2002 \& 2006; Mousis et al., 2002; Mosqueira \& Estrada, 2003;
Mousis \& Gautier, 2004; Alibert et al., 2005). Depending on the
precise structure of the subnebula, two scenarios can be invoked for
the formation of the regular satellites.  

 In this work, we only discuss the evolution of the gas phase
  within the Jovian subdisk, though the results are also likely to be
  applicable to the Saturnian system. In both scenarios discussed
  below, we follow Canup \& Ward (2002) in assuming the accretion of
  the satellites during large scale migration within the subdisk.

In the first scenario, solid particles which are dynamically coupled
to the gas (ranging in size from tens of centimeters to a few meters)
are accreted from the Solar nebula to the subnebula, which is
initially warm enough to vaporise them (or, at least, the ices they
contain). In this case, chemical reactions in the gas phase may lead
to the destruction/production of some volatile species (Prinn \&
Fegley, 1981 \& 1989). More importantly, high temperature gas phase
isotopic exchanges between molecular hydrogen and hydrogen-bearing
molecules may alter the deuterium concentration from that which was
acquired by the infalling material during its formation in the Solar
nebula. When the subnebula cools, the icy particles re-condense and
start to grow. As a result, the D:H ratios trapped within the icy
components of the planetesimals, and therefore within the regular
satellites themselves, are likely to be different from the original
values acquired in the Solar nebula.

In the second scenario, the regular satellites are formed later, from
a cold subnebula, in which the temperature is too low to cause the
vaporisation of the volatile components of infalling particles. The
regular satellites would therefore be expected to have D:H ratios
which reflect the Solar nebula at their epoch of formation, as
discussed in section 3.2.

Between these two extreme scenarios, a wide variety of setups are
possible, such that the regular satellites could have been formed from
a combination of re-vaporised and un-vaporised material.  In other
words, the two scenarios discussed represent the two end-members of a
continuum of models, through which the modified and un-modified
fractions of the material incorporated within the regular satellites
vary with the initial conditions. The following two sections discuss
these extremes in more detail, allowing us to draw conclusions on how
the measurement of the $f$-value within the regular satellites will
shed light on the details of the formation.

Note that the evolution and structure of the subnebula, as presented
above, is directly linked to the final stages of planetary
formation. In particular, a continuous flow of material from the solar
nebula to the subnebula occurs during the first epoch of the
subnebula's evolution. This inflowing material, which has a D:H value
which was acquired in the solar nebula, will therefore alter the
evolution of the deuterium enrichment in the subnebula. Such
considerations need to be taken into account when one attempts to
determine the formation of satellites in such a system. In the
diffusion equation presented below (Eq. \ref{diff}), a source term
would be needed to describe the vaporisation of these high-$f$
planetesimals.  

However, since it is difficult to formulate and incorporate
  such an expression in the diffusion equation, we did not consider
  the source term in this initial work. We intend to examine the
  problem in much more detail in the near future, and will address the
  influence of the vapourisation of high-$f$ planetesimals on the
  evolution of the global value of $f$ in the subnebula gas phase at
  that stage. Here, then, we will concentrate solely on the two
extreme closed subnebula models, described above, saving a more
precise calculation of D:H ratio evolution for future work.

\subsection{The formation of satellites in a warm subnebula}
\label{warm}

In this section, we consider the formation of regular satellites from
icy planetesimals produced in the Jovian subnebula. This hypothesis
implies that the Jovian subdisk was initially warm enough to vaporise
all the icy solids which formed in the solar nebula before being fed
into the disk. In order to provide a description of the thermodynamic
structure of the disk, we use the one-dimensional analytical turbulent
model developed by Dubrulle (1993) and Drouart et al. (1999).
 
\subsubsection{The Jovian subnebula model}

Our turbulent model of the Jovian subnebula is based on the work of
Shakura \& Sunyaev (1973), who characterise the turbulent viscosity
$\nu_t$ as

\begin{equation} 
\nu_t = \alpha \frac{C^2_S}{\Omega},
\end{equation}

where $C_S$ is the local sound velocity, $\Omega$ the Keplerian
rotation frequency and $\alpha$ the dimensionless viscosity
parameter. Since the physical origin of turbulence in accretion disks
is still not well established, this model is useful since it allows us
to describe the qualitative influence of whichever process is
responsible for the transport of angular momentum through the
disk. Our model also incorporates the opacity law developed by Ruden
\& Pollack (1991; see e.g. Drouart et al., 1999 for details). The
temporal evolution of the disk temperature, pressure and surface
density profiles depends upon the evolution of the accretion rate,
$\dot{M}$, which we define (following Makalkin \& Dorofeeva (1991) to
be:

\begin{equation} 
\dot{M} = \dot{M}_0 (1 + \frac{t}{t_0})^{-s}.
\end{equation}

$\dot{M}$ decreases with time following a power law which is determined
by the initial accretion rate $\dot{M}_0$ and the accretion timescale
$t_0$. In this work, we adopt $s$ = 1.5, a value that allows our law to be
consistent with that derived from the evolution of accretion rates in
circumstellar disks (Hartmann et al., 1998). The accretion timescale
$t_0$ is calculated by Makalkin \& Dorofeeva (1991) to be:

\begin{equation} 
t_0 = \frac{R^2_D}{3 \nu_D},
\end{equation}

where $\nu_D$ is the turbulent viscosity at the initial radius of the
subdisk, $R_D$. Three parameters constrain $\dot{M}_0$ and $t_0$: the
initial mass of the disk $M_{D0}$, the coefficient of turbulent
viscosity $\alpha$ and the radius of the subnebula $R_D$.

\subsubsection{Thermodynamic conditions within the subnebula}

Table \ref{parasub} summarises the thermodynamic parameters used in
our turbulent model of the Jovian subnebula. The outer radius of the
subnebula is taken as 150 $R_J$ $\sim$1/5 $\times$ $R_{Hill}$, a
value close to that calculated by Magni \& Coradini (2004) using a 3D
hydrodynamical model of the final stages of Jupiter's formation. The
initial accretion rate chosen in our model is a high value, namely
$10^{-5} \Msun / yr$, allowing us to describe a highly viscous disk in
which temperature and pressure conditions are elevated, particularly
at early epochs. The final parameter, $\alpha$, which governs the
viscosity, is, at best, poorly constrained by observational
data. However, in an attempt to reproduce the ice/rock fractions of
the Jovian satellites, Alibert et al. (2005) derived a range of
possible values for the viscosity parameter. In this model, we use
their favoured value, namely $2 \times 10^{-4}$.

\begin{figure} 
\resizebox{\hsize}{!}{\includegraphics[angle=-90]{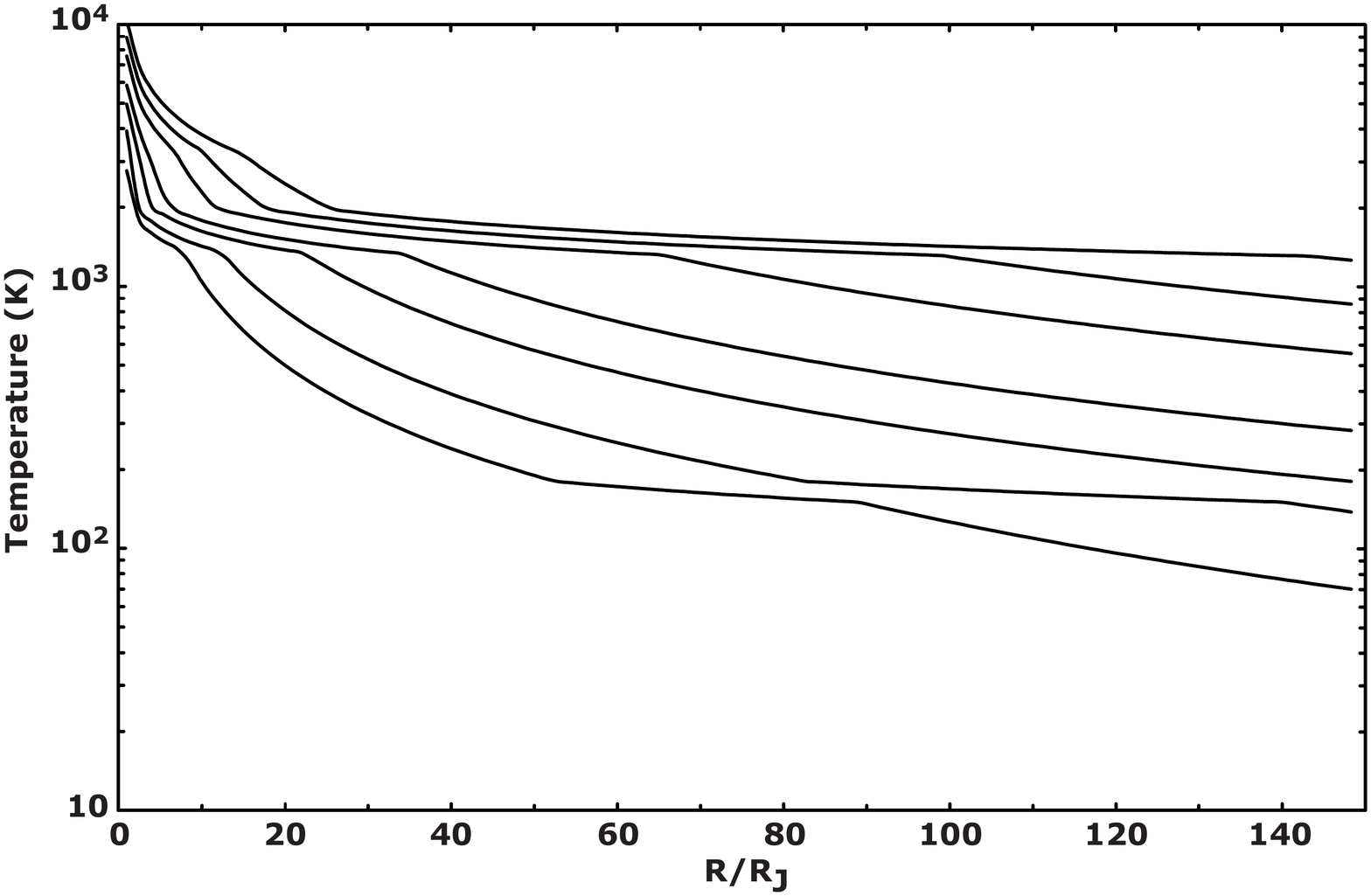}}
\caption{Temperature profiles at different epochs in the midplane of the Jovian subnebula, at times (from top to bottom) $t$ = 0, 100 yr, 300 yr, 10$^3$ yr, 2$\times$10$^3$ yr, 5$\times$10$^3$ yr and 10$^4$ yr.}
\label{Tsub} 
\end{figure}

\begin{figure} 
\resizebox{\hsize}{!}{\includegraphics[angle=-90]{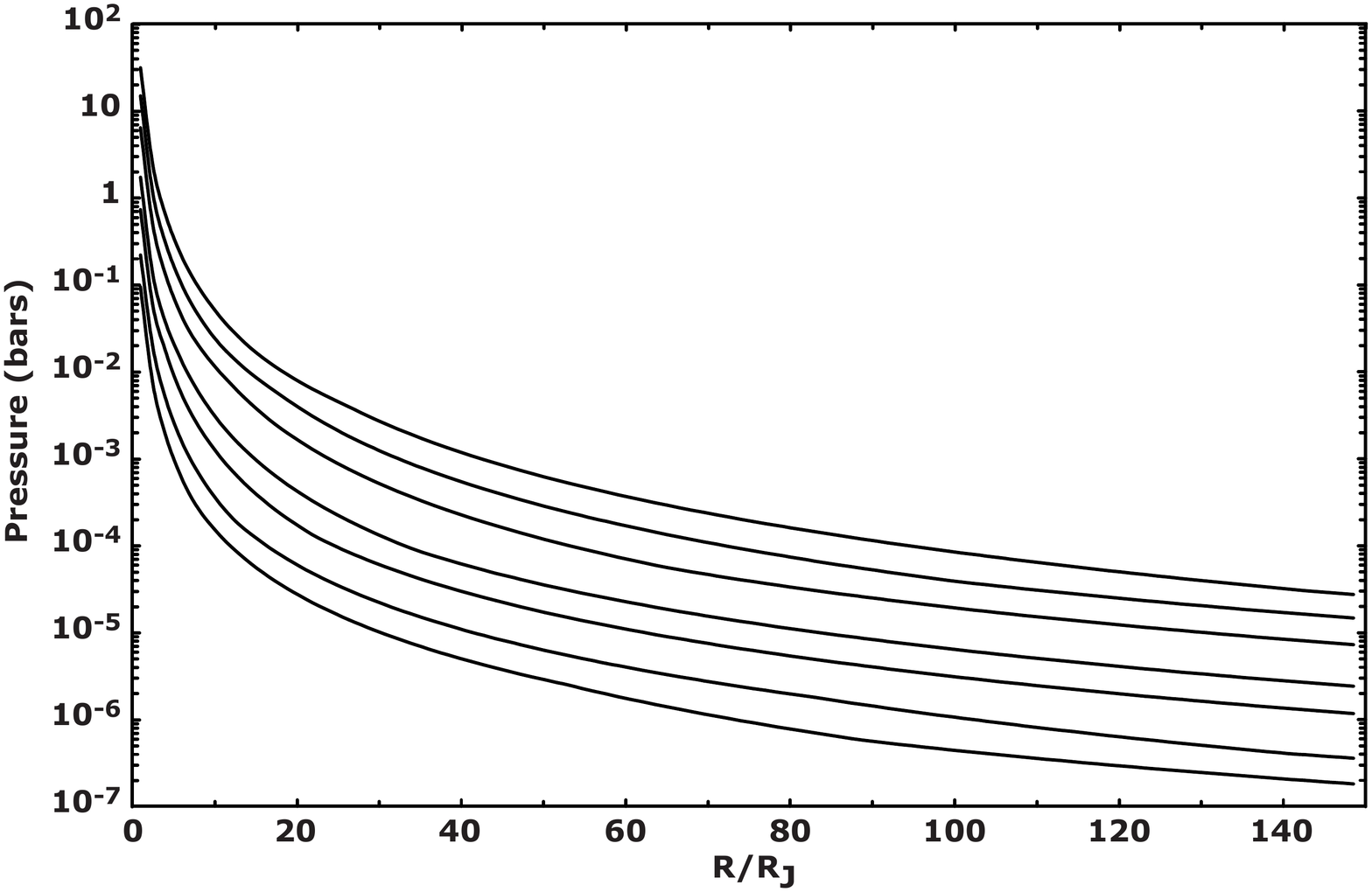}}
\caption{Pressure profiles at different epochs in the midplane of the Jovian subnebula. Times are the same as in Fig. \ref{Tsub}.}
\label{Psub} 
\end{figure}

\begin{figure} 
\resizebox{\hsize}{!}{\includegraphics[angle=-90]{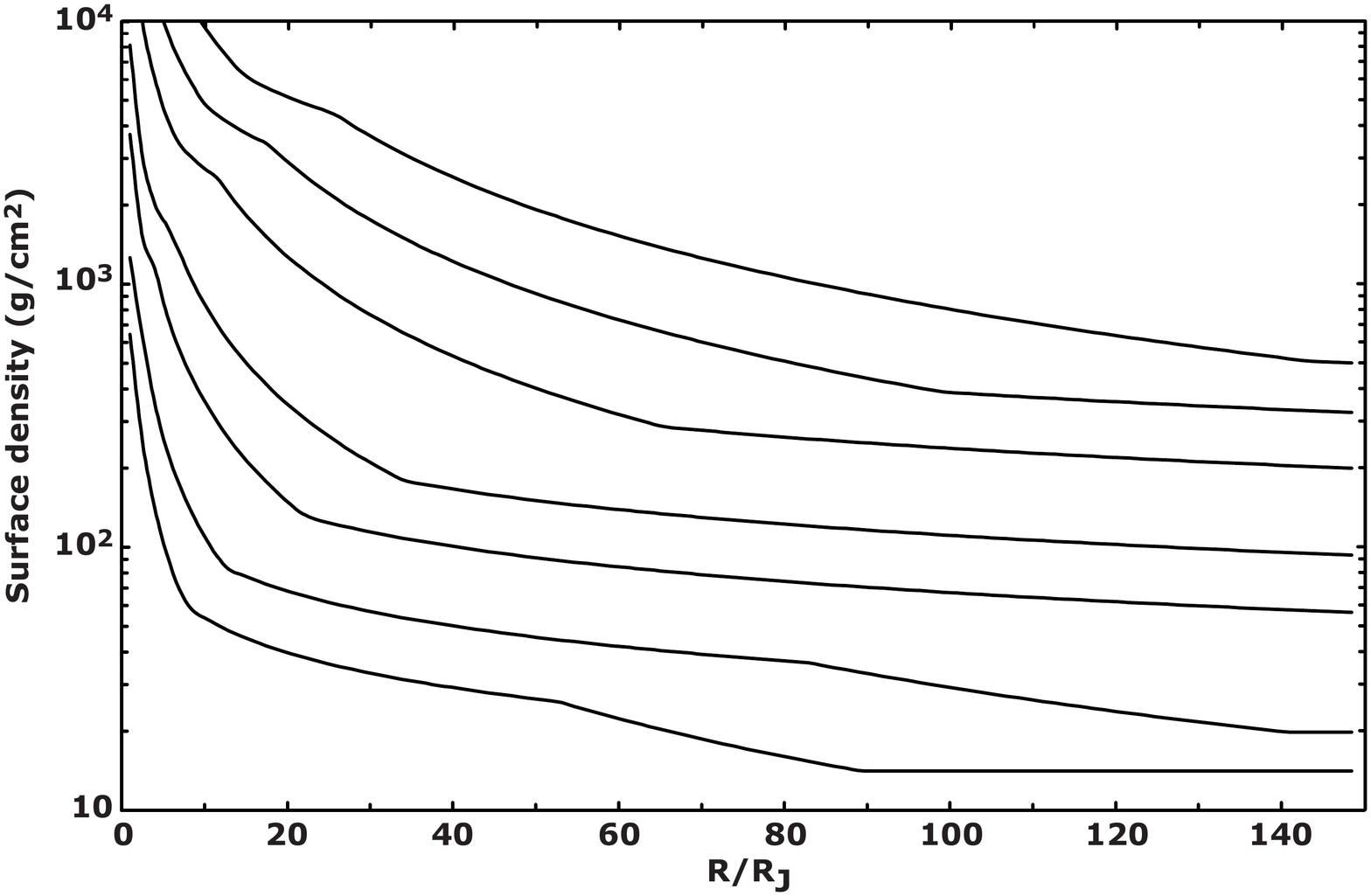}}
\caption{Surface density profiles at different epochs in the midplane of the Jovian subnebula. Times are the same as in Fig. \ref{Tsub}.}
\label{Ssub} 
\end{figure}

Figures \ref{Tsub}--\ref{Ssub} show radial profiles of temperature
$T$, pressure $P$ and surface density $\Sigma$, respectively, at
various epochs during the evolution of the subnebula. Water is in the
vapour phase at $t$ = 0 throughout the subnebula, and $T$, $P$, and
$\Sigma$ decrease over time and as a function of the distance to
Jupiter. At $t$ $\sim$4000 yr, water starts to crystallise at the
outer edge of the subdisk. The cooling of the subnebula results in the
inward migration of the water condensation front, which reaches the
orbits of Callisto (26.6 $R_J$), Ganymede (15.1 $R_J$), Europa (9.5
$R_J$) and Io (6 $R_J$) at $t$ = 0.06 Myr, 0.14 Myr, 0.27 Myr and 0.6
Myr, respectively.

\subsubsection{Isotopic exchange}

In the warm Jovian subnebula, the evolution of $f$ is governed by the
equation of diffusion previously detailed by Drouart et al. (1999) and
Mousis et al. (2000), which is:

\begin{equation}
\partial_tf=k(T)P(A(T)-f)+\frac{1}{\Sigma R} \partial_R (\kappa R \Sigma \partial_R f).
\label{diff}
\end{equation}

The first term in the right-hand side of Eq. \ref{diff} describes 
isotopic exchange between HDO and H$_2$. Function $A(T)$ is the
isotopic fractionation at equilibrium, $k(T)$ is the rate of the
isotopic exchange between HDO and H$_2$, and $P$ is the total
pressure. $A(T)$ is derived from the tables given in Richet et
al. (1977), and the rate $k(T)$ is taken from experiments carried out 
by L{\' e}cluse~and~Robert~(1994). The second term on
the right-hand side of Eq. \ref{diff} describes turbulent diffusion
throughout the subnebula. This term is a function of the 
local surface density $\Sigma(R,t)$, and on the diffusivity $\kappa$. Following
Drouart et al. (1999), $\kappa$ is assumed to be the ratio of the
turbulent viscosity to the Prandtl number $P_R$, the value of which is always
close to the unity.

As in the case of the Solar nebula, Eq. \ref{diff} is valid as long as
H$_2$O does not condense in the Jovian subdisk. This implies that the
final value of deuterium enrichment in the microscopic ices
incorporated within the regular satellites is that obtained at the
time and at location at which the vapour condenses. As soon as the
grains become millimeter-sized, they begin to decouple from the gas,
and continue to grow, eventually forming planetesimals.  Whatever the
subsequent evolution of these solids, however, the D/H ratio within
them will be that of these microscopic grains.

\begin{figure} 
\resizebox{\hsize}{!}{\includegraphics[angle=-90]{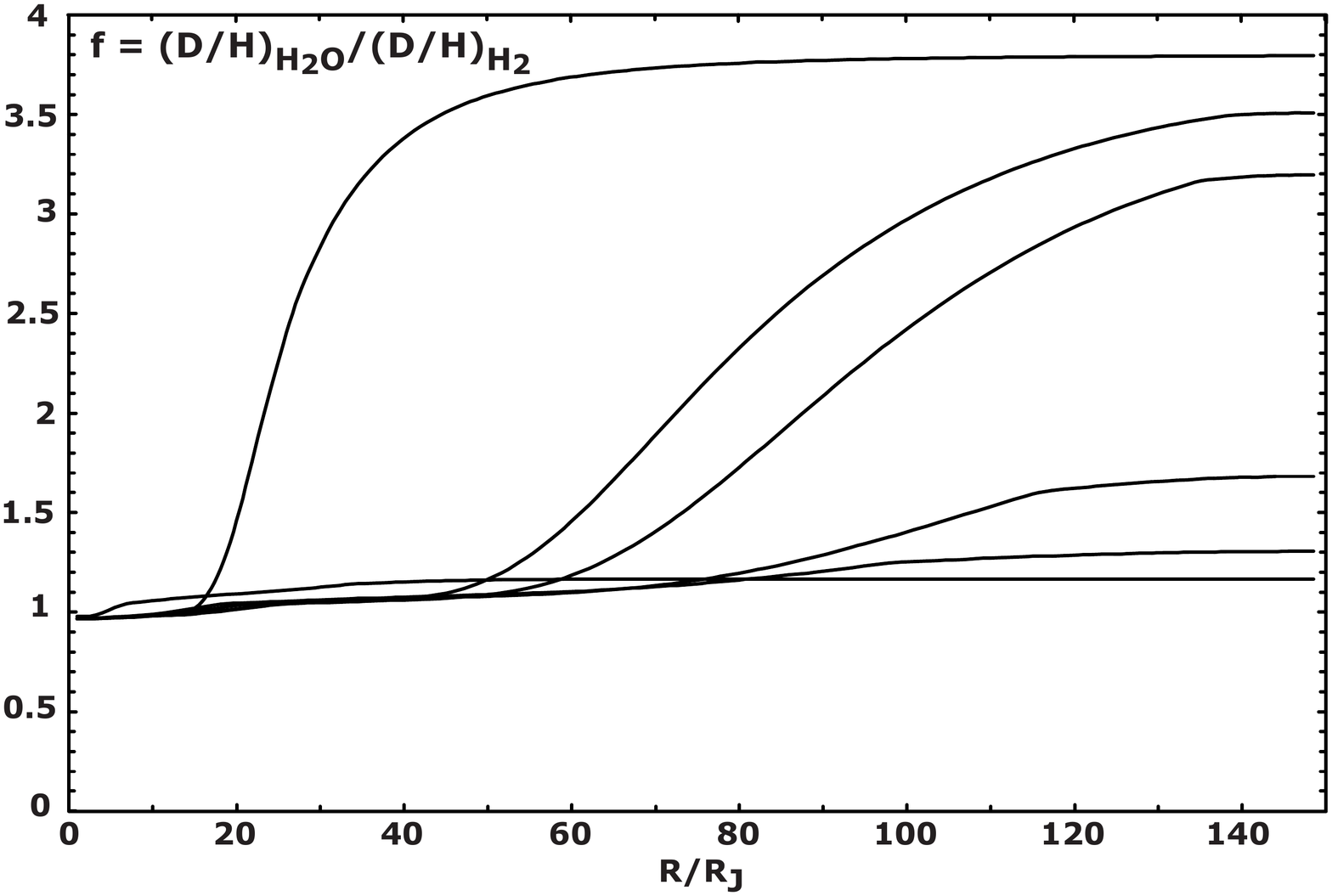}}
\caption{Enrichment factor $f$ of the D:H ratio in H$_2$O with respect to the protosolar value in the Jovian subnebula midplane, as a function of the distance to Jupiter, at times (from top to bottom) $t$ = 1 yr, 5 yr, 10 yr, 50 yr, 100 yr and 10$^3$ yr. Calculations are made for the deuterium exchange between water and hydrogen in the vapor phase. They are stopped when water is condensed, a process that occurs closer and closer to Jupiter when time increases. The value for $f$ at $t$ = 0 is taken to be equal to 3.8 (the minimum deuterium enrichment value in water ice condensed at 5.2 AU in the solar nebula -- see e.g. Fig. \ref{DHsol}), whatever the distance to Jupiter in the subdisk.}
\label{DHsub} 
\end{figure}

The enrichment factor, $f(R,t)$, can then be obtained by integrating
Eq. \ref{diff}, which requires the determination of spatial and
temporal boundary conditions. The spatial boundary conditions
are obtained by setting $\partial f/\partial R$~=~0 at both
$R~=~1~R_J$ and $R~=~R_D$. When condensation occurs within $R_D$, we
set $\partial f/\partial R$ = 0 at the radius of condensation. Since
the vapor phase of the Jovian subnebula initially consisted of both
gas and vaporized ices falling in from the Solar nebula, the initial
D:H ratio in water vapor within the subdisk is that of the water ice
produced in the feeding zone of Jupiter. As mentioned in
Sec. \ref{background}, the value of $f$ in H$_2$O ice produced in the
feeding zone zone of Jupiter is between 3.8 and 4.7 times the Solar
value. Setting $f(R)$~=~3.8 at $t$~=~0 when integrating Eq. \ref{diff}
reveals that, as shown on Fig. \ref{DHsub}, $f(R,t)$ rapidly decreases
and reaches values lower than $\sim$1.2 throughout the entire
gas phase of the Jovian subnebula. Setting $f(R)$~=~4.7 at $t$~=~0
when integrating Eq. \ref{diff} leads to the same result. Once this
material has condensed within the formation zone of the regular
satellites, at late epochs, the water ice will keep this low deuterium
enrichment value. 

The value reaches $f$~$\sim$1 very rapidly, well before the
water condenses, at which point a stable equilibrium has been reached
between the two deuterium reservoirs (H$_2$ and H$_2$O). As a result, regular icy satellites accreted from icy planetesimals produced $in~situ$ will also retain this value of $f$ in
their water ice.

Note that CO to CH$_4$ and CO$_2$ to CH$_4$ gas phase conversions may
occur within a warm and dense Jovian subdisk via the following
reversible reactions (Prinn \& Fegley, 1981; Mousis \& Alibert, 2006):

\begin{equation}
\rm CO + 3H_2 = CH_4 + H_2O,
\end{equation}

\begin{equation}
\rm CO_2 + 3H_2 = CH_4 + 2H_2O.
\end{equation}

Such gas phase reactions would lead to the production of additional
water in the Jovian subnebula from hydrogen with a Solar D:H
ratio. Since the D:H ratio in the water produced via these reactions
would then exhibit Solar D:H, our conclusion that the regular icy
satellites would exhibit $f$ of order unity would not be affected.

Finally, it should be noted that, although the calculations shown above have been
carried out in the context of the Jovian subnebula, they would be equally valid within 
the Saturnian disk.

\subsection{The formation of satellites in a cold subnebula}

In the previous sections, we have considered the case where material
entering a planetary subnebula would be vaporised and undergo a period
during which further reactions are possible. 
In the opposite scenario, however, where the planetary subnebula is sufficiently cold that the
material is never revaporised, the situation is far simpler. The
material which is used to form the regular satellites of the planet
would be representative of that making up the subnebula, which would,
in turn, be the same as that in the general vicinity of the planet
within the Solar nebula. As a result, one would expect that, for the
``cold subnebula'' hypothesis, the regular planetary satellites would
exhibit D:H ratios identical to that present in the Solar nebula at
the heliocentric formation distance of their parent planet.

However, should that planet has migrated a significant distance
between its formation and the formation of its satellites, we would
expect that the satellites would display $f$-values representative of
the regions through which the planet migrated as they formed. Hence,
if some satellites formed earlier in the migration than others, they
could display some variation in their incorporated deuterium content,
and individual satellites may have values which result from the
combination of deuterated ice from various locations in the planetary
migration. However, unless the planets migrated over a very large
distance, these variations should be fairly minor.

It is clear, therefore, that there would be expected to be a
significant difference in the measured D:H values for the regular icy
planetary satellites between these two models (the ``hot'' and
``cold'' subnebula approaches). This means that measurements of the
D:H value in these satellites are critical in helping our
understanding of subnebula processes and the general conditions
present at the time of the formation of the regular icy satellites.

 At a first glance, it appears that the idea of a ``cold''
  subnebula is, at the very least, at odds with the volatile-poor
  compositions of both Io and Europa. Indeed, it could be argued that
  these bodies argue for a ``hot'' subnebula - at least in the inner
  regions closest to Jupiter. However, when one considers the temporal
  evolution of the subnebula, that is, the way in which the subnebula
  cools over time, it is quite possible that the two innermost
  Galilean satellites formed early, while the nebula was still hot
  enough to vaporise the volatile content of inbound planetesimals,
  whilst the outer moons would form later, after the nebula has cooled
  sufficiently that the remaining infall of planetesimal material
  remains frozen throughout. This ``hybrid'' model has previously been
  discussed (e.g. Mousis \& Gautier 2004, and Alibert et
  al. 2005). In other words, it is possible that the various
  satellites formed in different thermodynamic regimes, typified by
  our two extreme cases.

\section{Deuterium enrichment in the irregular satellites and beyond}

Beyond the case of the regular planetary satellites, the study of
deuterium could offer insights into the formation mechanisms for other
populations of icy bodies. Horner et al. (2007) made a study of the
way that $f$ measured in icy planetesimals (particularly the comets we
observe today) would be affected by their formation location in the
Solar nebula. It is clear that the application of these ideas need not
be limited to the study of cometary bodies, and so we here extend our
arguments to the other icy bodies in the outer Solar system, an ever
increasing number of which have been discovered in recent years. One
family of icy bodies whose numbers have grown rapidly in the last
decade are the irregular satellites of the outer planets. Ongoing
surveys (e.g. Sheppard et al., 2006) have now found many such bodies
orbiting around Jupiter, Saturn, Uranus, and Neptune, with a roughly
equivalent number in each case. This is a somewhat unexpected result
when one considers the huge range of semi-major axes and mass that is
covered from Jupiter to Neptune. The satellites themselves seem to
gather in discrete dynamical families - which suggests that a much
smaller population of objects has slowly been collisionally shattered
to provide that we see today. The satellites orbit far further from
their parent planets than the regular satellites discussed earlier,
with a much wider range of eccentricities and inclinations. Indeed,
many of the irregular satellites orbit their parent planets in a
retrograde fashion. These orbits are believed to the result of the
parent bodies of the irregular families having been captured during
the final stages of planet formation. 

Whilst the catalogue of icy Solar system objects has been growing
rapidly, a number of authors have been looking into methods by which
these various populations of object could be formed (e.g. Morbidelli
et al., 2005, on the permenent Jovian Trojans; Horner \& Evans, 2006,
on the capture of Centaurs onto Trojan orbits and Jewitt et al.,
2007 on the capture of the irregular satellites). From the various
formation mechanisms proposed for the different reservoirs, it is
clear that these bodies could have formed from material sourced from
throughout the outer Solar system, and hence would be expected to
display $f$-values significantly different to those expected from
objects which formed at 5 Au. Indeed, the first measurements
  of the mass and density of a binary object within the Jovian Trojan
  population (Patroclus, found to have $\rho \sim 0.8 \pm 0.15 g/cm^3$
  (Marchis et al., 2006)) have shown that at least some of these
  objects are similar in nature to objects in the outer Solar system
  (binary Edgeworth-Kuiper belt objects, for example), rather than
  being dense rocky or metallic objects.

Given that these families of icy bodies are likely to be
captured members of these parent populations, we can apply the same
arguments to them as applied by Horner et al. (2007) to the
comets. Studies of $f$ in the ice of a particular irregular satellite,
for example, could then give information on its formation region
within the Solar nebula - with chaotic scattering of the icy bodies in
the disk, it is quite possible that the irregular satellites formed
throughout the outer Solar system, and thus that they will exhibit a
wide range of $f$-values. That said, if the bodies within a family of
irregular satellites truly represent the remains of a single,
collisionally fractured parent body, then we would expect that the
$f$-value measured for all objects within a family would be the
same. Different families would have a different $f$, reflecting their
different formation regions, but all bodies in given family would have
the same $f$. It is also clear that, given their different formation
mechanism, the irregular satellites should display $f$-values that are
greatly different to those that would be observed in the regular
family - not only did they likely form at different heliocentric
distances, but the material from which they formed was also not
reprocessed in the planetary subnebula, and as such would be expected
to contain significantly more deuterium within their water than is
present in the regular satellites.

Next, consider the various theories of Jovian Trojan formation. From
the above, it is clear that, had the Trojan asteroids formed {\it in
  situ}, then they would display $f$-values typical of objects forming
at $\sim$5 AU from the Sun. If, however, the Trojans were captured
late in the development of the Solar system (for example, during the
Late Heavy Bombardment, e.g. Morbidelli et al., 2005), it is possible
that the bulk of the material which formed between the orbits of
Jupiter and Neptune would have been cleared, and thus that the Trojans
could primarily have originated in the youthful Edgeworth-Kuiper
belt. This would, in turn, mean that the bulk of the objects would
display similar, high values of $f$. Had the Trojans instead been
captured at an earlier epoch, one might expect them to display a far
wider range of $f$-values, representative of those for objects forming
in the whole outer Solar system. 

\section{The future of extra-terrestrial deuterium enrichment measurements}

Observations of the D:H ratio in water have been carried out for the
terrestrial planets (e.g. Encrenaz et al., 1995, Makrides et al., 2006
etc.), together with comets Hyakutake (Bockel{\'e}e-Morvan et al.,
1998), Hale-Bopp (Meier et al., 1998) and Halley (Balsiger et al.,
1995). Additionally, the D:H ratio in molecular hydrogen has been
measured for the giant planets (e.g. Feuchtgruber et al., 1999,
Lellouch et al., 2001), while the value of the D:H ratio within
CH$_4$:CH$_3$D in the atmosphere of Titan has been measured from the
Earth (e.g. Coustenis et al., 2003) and by Cassini (e.g. B{\'e}zard et
al., 2007). However, beyond this restricted set of measurements,
little is known about the amount of deuterium incorporated in the
bodies of our Solar system. Fortunately, in the forthcoming years, a
number of new instruments should shed new light on the deuteration of
the Solar system.

Since they cover the far infrared/millimeter range, the Herschel Space
Observatory and Atacama Large Millimeter Array (ALMA) will both
provide an opportunity for observers to measure $f$ in the atmospheres
of the planets, and comae of comets. We will have access to the
Centaurs and short-period comets, and any object that could exhibit a
tenuous atmosphere. It is now well understood that the
  short-period comets are the daughter population of the Centaurs
  (e.g. Horner et al., 2004). Therefore, it is clear that, even if no
  Centaur is active enough to be studied with these instruments,
  additional measurements of deuterium in such comets will provide new
  constraints on the range of $f$-values which exist in these
  bodies (as discussed in Horner et al., 2007). In turn, this will allow us to place better constraints on
the formation regions of the various cometary reservoirs. For a more
detailed review of the measurements which will be possible with these
instruments, we direct the reader to Encrenaz et al. (2005).

 Rosetta will rendezvous with comet 67P/Churyumov-Gerasimenko in
  2014. Onboard the main spacecraft is a near-infrared spectrometer,
  VIRTIS, that will provide spectra in the near infrared (2-5 microns)
  at spectral resolutions R between 300 and 1000. Detection of
  deuterated species such as HDO and CH$_3$D on the nucleus will be
  possible (Coradini et al., 1998). The MODULUS Ptolemy experiment on
  the Roland lander will obtain isotopic ratios of the major volatiles
  \emph{in situ} by ion trap mass spectrometry and gas chromatography
  (Biele et al. 2002). These measurements, offering us the
    chance to look at a Jupiter family comet close up, will allow us
    further information on the $f$-value in Jupiter-family comets, and
    their parents, the Centaurs.

The James Webb Space Telescope (hereafter JWST) will allow the D:H
ratio in many solar system objects to be probed beginning in
2014. This will be possible thanks to two instruments, the Mid-IR
Instrument (MIRI) and the multi-object spectrometer (NIRspec). MIRI
will provide broadband field imagery and medium resolution
spectroscopy (up to R = 3000) between 5 and 27 microns (29 microns for
the spectroscopy), while NIRspec will allow simultaneous spectroscopic
observations at comparable spectral resolution, of up to 100 objects
at wavelengths between 0.6 and 5 microns.  It will be possible to
examine the inner comae of active comets with NIRSpec, allowing high
precision measurements of the degree of deuterium enrichment. For the
largest and closest KBOs, it will also be possible to obtain spectra,
allowing the identification of major molecules and isotope ratios
(including deuterated species) for objects such as Triton, Pluto,
Quaoar and Varuna (Gardner et al., 2006). It should therefore also be
possible to measure the value of D:H in the water ice on the surface
of the Jovian satellites, together with taking advantage of
  the outgassing from Enceladus to measure the $f$-value for the
  Saturnian satellite in some detail.

It has been suggested that the dust clouds recently discovered in the
vicinity of the Galilean satellites in the Jovian system are the
result of hypervelocity impacts of interplanetary micrometeorites upon
their surface (e.g. Krivov et al. 2002; Kr{\"u}ger et al. 2003,
2006). Collection of this dust by a spacecraft would therefore enable
us to study the $f$-value present in the ices of these worlds. For
this reason, looking still further into the future, it is vital that
the proposed Laplace mission, scheduled to launch in 2017 at the
earliest, includes instruments dedicated to the analysis of both the
chemical and isotopic composition of dust particles collected while in
the Jovian system. In particular, we believe that the mission should
make a priority of the collection of dust in the vicinity of the icy
satellites during close fly-bys. In order that the value of $f$ in
these grains can be measured, the mission will need to incorporate a
dust collector with an ablation system which would vaporise them. A
high resolution mass spectrometer would then allow the measurement of
D:H in hydrogen bearing volatiles. It is important to note that such
an instrument would need to have a significantly higher resolution
than the Ion and Neutral Mass Spectrometer on Cassini. That
instrument, although it has obtained many exciting results, sadly
lacks the resolution necessary to separate different species
  with the same atomic weight, having a resolution of just 1
  amu. Therefore, the identification of deuterated species is not
  possible due to the presence of other molecules of the same weight
  (e.g. HD$^{16}$O vs. H$_2$$^{17}$O; NH$_3$ vs. CH$_3$D).

With regard to the irregular satellites, it may be that a close
approach by a mission (such as Laplace) en-route to Jupiter may reveal
traces of outgassing from one of these icy bodies, allowing a direct
measurement of the $f$-value within the body to be made.  
However, it is more likely that such a close approach would allow such
a mission to collect dust and ice sputtered from the surface of these
satellites in the same way as with the regular satellites.

We suggest that such a flyby would be of great value to the scientific community
as a whole, and hope that the proposed missions can find room in their
schedules for a visit to the irregular regime.

\section{Summary and discussion}

In previous work (Horner et al., 2007), we examined the role that
variations in the D:H ratio through the Solar nebula would have on
cometary objects observed to originate from different reservoirs in
the outer Solar system. The goal of that work was to highlight that
observations of the D:H ratio in such objects could prove a
cornerstone in helping our understanding of the different regions in
which the various cometary populations formed.

Here, we have extended our arguments to include the other populations
of hydrated objects in the outer Solar system - the icy planetary
satellites (both regular and irregular), and the Jovian and Neptunian
Trojans. We have shown that, since these objects cover a wide range of
formation scenarios, the study of the D:H incorporated in their water
ice provides a useful tool to answer questions on their origin. Did
the regular satellites form in a hot or cold subnebula?  Do the
irregular planetary satellites truly represent a captured and then
shattered population of objects? If so - from where were they
captured? What was the origin of the Jovian and Neptunian Trojans?

We focus, in particular, on the formation of regular satellites in a
circum-planetary disk of gas and dust around Jupiter -- the Jovian
subnebula. Current theories which describe the formation of such
satellites cover a wide range of initial conditions. However, the two
extreme cases are the ``hot'' and ``cold'' subnebula models. In the
former, the entire subdisk is sufficiently warm that icy material
falling into the subnebula from the Solar nebula is entirely
vaporised, allowing the exchange of deuterium between molecular
hydrogen and water to continue in the gas phase, when the water
involved would otherwise already be trapped as ice. The other extreme,
the ``cold'' model, assumes that the subnebula was sufficiently cold
that none of the infalling volatiles were vaporised prior to their
accretion in the planetesimals which went on to form the
satellites. Although the true formation scenario for the regular
satellites undoubtably lies somewhere between these extremes, they
allow us to constrain the behaviour of the ices which went on to form
the satellites we currently observe. In the case of the ``hot''
subnebula model, we have shown that the effect of the re-vaporisation
of the infalling water ice allows gas phase reactions between HDO and
$H_2$ to occur, leading to the gradual depletion of deuterium within
the initially D-rich water. As a result, satellites which formed in
this way would be expected to exhibit D:H ratios in their ice which
are close to the Solar value. By contrast, in the case where the
satellites form in a ``cold'' subnebula, since vaporisation of the
volatiles does not occur, the D:H ratio within the satellite ices
would be significantly higher, representative of material within the
Solar nebula at the heliocentric distance at which the satellite's
parent planet formed. The measurement of the D:H ratios within the
regular satellites therefore provides a key constraint for the
discussion of their formation.  Although we do not go into specific
detail, the models described above are equally applicable for
satellites forming in the Saturnian subnebula.

In the future, new observatories and space missions will allow us to
measure $f$ with a greater accuracy, and in many more objects, than
ever before.  HERSCHEL, ALMA, Rosetta, and the JWST will allow the
measurement of deuterium in objects as diverse as KBOs, comets, and
the satelites of the giant planets, while the proposed Laplace Jupiter
orbiter offers a unique opportunity to measure deuterium {\it
  in-situ}, through use of a dust collector/ablation system, allowing
the direct comparison of the regular and irregular Jovian satellites,
and offering new insights into their formation.  Early discussion of
the type of measurements required is vital for those involved in the
planning stages of these projects, in order that appropriate
instrumention is constructed to make the required measurements.

Another factor which may have influenced the final value of $f$
obtained by objects (particularly the giant planets and their regular
satellites) is that the giant planets are thought to have migrated
over a significant distance during their formation and subsequent
evolution. In the most extreme examples, it has even been suggested
that Uranus and Neptune formed between the orbits of Jupiter and
Saturn (Levison et al., 2004)!  In these cases, it is likely that the
value of $f$ shown by the regular satellites of these planets will
provide a doubly useful tool in determining the true history of the
outer Solar system, even though it seems likely that these satellites
have been destroyed and reassembled since their formation
(e.g. Banfield \& Murray, 1992). If it is the case that Uranus and
Neptune formed far closer to the Sun than their current location, then
it seems likely that the material incorporated in their regular
satellites could reflect their formation and migration. It must be
pointed out, here, that current measurements of the $f$-value
  in the giant planets themselves provides no useful constraint on
  their formation location, due to the fact that the measured value is
  heavily affected by the gaseous hydrogen incorporated during their
  formation, in addition to water obtained from planetesimals. Given a
  sufficiently good interior model (e.g. Feuchtgruber et al. 1999), it
  may be possible to extract the native $f$-value of the accreted icy
  fraction of the planet, which would clearly prove very useful in the
  study of the planetary migration. However, the satellites currently
  offer a simpler solution to the problem, since they formed solely
  from the accretion of planetesimals.  Studies of these objects could
  provide detailed information on their $f$-values, which can
constrain their migration and formation histories. Were the satellites
present prior to the migration? Did they form afterward, or during the
process?  Similarly, should it turn out that the ``cold subnebula''
model is the fairest representation of the formation of the regular
Jovian and Saturnian satellites, then it is possible that these
objects contain, trapped within their ices, a record of the migration
of their parent planets through the Solar nebula. Clearly, such
measurements could even be used to place constraints on the distance
over which the planets migrated.

\section*{Acknowledgements}

This work was supported in part by the French Centre National d'Etudes Spatiales. JH gratefully acknowledges the financial support provided by STFC. We thank Torrence Johnson and Emmanuel Lellouch for their helpful remarks.

\clearpage

\begin{table}
\caption{Initial radius R$_{D0}$, mass M$_{D0}$ and viscosity parameter $\alpha$ of the minimum-mass and maximum-mass solar nebula models used by Mousis (2004).}
\begin{center}
\begin{tabular}[]{lccc}
\hline
\hline
\noalign{\smallskip}
		& R$_{D0}$ (AU)	& M$_{D0}$ (\Msun)	& 	$\alpha$ \\
\noalign{\smallskip}
\hline
\noalign{\smallskip}
Maximum mass solar nebula 	&	27	&	0.3	&	0.003 \\
Minimum mass solar nebula 	&	15	&	0.06	&	0.003 \\
\noalign{\smallskip}
\hline
\end{tabular}
\end{center}
\label{paraneb}
\end{table}
\clearpage

\begin{table}
\caption{Thermodynamic parameters adopted for the warm subnebula.}
\begin{center}
\begin{tabular}[]{lc}
\hline
\hline
\noalign{\smallskip}
Thermodynamic parameters		&	 \\
\noalign{\smallskip}
\hline
\noalign{\smallskip}
Mean mol. weight (g/mole)		& 	2.4	 \\
$\alpha$ 				& 	2 $\times$ 10$^{-4}$ \\
Disk's radius ($R_J$) 			& 	150		    \\
Initial disk's mass ($M_J$)		& 	3 $\times$ 10$^{-3}$ \\
Initial accretion rate ($M_J$/yr)	& 	1 $\times$ 10$^{-5}$ \\
Accretion timescale (yr)		& 	140 \\
\noalign{\smallskip}
\hline
\end{tabular}
\end{center}
\label{parasub}
\end{table}
\clearpage

\clearpage

\clearpage

\clearpage

\clearpage

\clearpage

\end{document}